\documentclass[epj]{webofc}
\usepackage[utf8]{inputenc}
\usepackage[varg]{txfonts}   
\usepackage{booktabs}
\usepackage{xcolor}
\definecolor{darkred}{rgb}{0.4,0.0,0.0}
\definecolor{darkgreen}{rgb}{0.0,0.4,0.0}
\definecolor{darkblue}{rgb}{0.0,0.0,0.4}
\usepackage[bookmarks,linktocpage,colorlinks,
    linkcolor = darkred,
    urlcolor  = darkblue,
    citecolor = darkgreen]{hyperref}
\usepackage{subfigure}
\usepackage{units}
\wocname{EPJ Web of Conferences}
\woctitle{Lattice2017}
\begin{document}
%
\selectlanguage{english}
\title{%
The pion quasiparticle in the low-temperature phase of QCD
}
\author{%
\firstname{Bastian B.} \lastname{Brandt}\inst{1} \and
\firstname{Anthony} \lastname{Francis}\inst{2} \and
\firstname{Harvey B.} \lastname{Meyer}\inst{3} \and
\firstname{Daniel} \lastname{Robaina}\inst{4} \and
\firstname{Kai} \lastname{Zapp}\inst{3}\fnsep\thanks{Speaker, \email{kazapp@uni-mainz.de} }
}
\institute{%
Institut f\"ur Theoretische Physik, Johann Wolfgang Goethe-Universit\"at, \\ D-60438 Frankfurt am Main, Germany
\and
Department of Physics \& Astronomy, York University, Toronto ON M3J 1P3, Canada
\and
PRISMA Cluster of Excellence, Institut f\"ur Kernphysik and Helmholtz-Institut Mainz, \\Johannes Gutenberg-Universit\"at Mainz, D-55099 Mainz, Germany
\and 
Institut f\"ur Kernphysik, Technische Universit\"at Darmstadt, D-64289 Darmstadt, Germany
}
\abstract{%
 We extend our previous studies [PhysRevD.90.054509, PhysRevD.92.094510] of the pion quasiparticle in the low-temperature phase of two-flavor QCD with support from chiral effective theory. This includes the analysis performed on a finite temperature ensemble of size $20\times 64^3$ 
at $T\approx 151$MeV and a lighter zero-temperature pion mass $m_{\pi} \approx 185$ MeV. Furthermore, we investigate the Gell-Mann--Oakes-Renner relation at finite temperature and the Dey-Eletsky-Ioffe mixing theorem at finite quark mass.
}
\maketitle
\section{Introduction}\label{intro}
The study of strongly interacting matter under extreme conditions such as finite temperature
has presented a theoretical and experimental challenge for many years. In experiment, 
heavy-ion collisions offer the possibility to study hot and dense QCD matter under laboratory conditions. One objective is to gain insights into the deconfined phase, namely the Quark-Gluon-Plasma, where the fundamental degrees of freedom of QCD, i.e., quarks and gluons, are expected to become quasiparticles. Therefore it is important to investigate how the zero-temperature excitations get modified with increasing temperature. On the theory side, often the Hadron Resonance Gas Model is used in the low-temperature phase. 
It describes the thermodynamics by assuming the medium to consist of a non-interacting gas of hadrons and resonances up to a cut-off mass. However, its success in providing estimates for equilibrium properties of the medium, such as quark susceptibilities, does not imply that individual excitations of the spectrum are in any sense similar to their vacuum analogues. 

Here we present an extension of our study~\cite{Brandt:2014qqa, Brandt:2015sxa} concerning the pion quasiparticle in the low-temperature phase of two-flavor QCD. In this article, we test the modified dispersion relation of the pion quasiparticle on a new ensemble with a lighter zero-temperature pion mass. Furthermore, we investigate the Gell-Mann--Oakes--Renner relation at finite temperature as well as the Dey-Eletsky-Ioffe mixing theorem \citep{Dey:1990ba} at finite quark mass. 

\section{Dispersion relation of the pion quasiparticle} 
\label{sec-Disp}
In this section we describe our analysis of the modified dispersion relation of the pion quasiparticle as it is proposed by thermal chiral perturbation theory~\cite{Son:2001ff,Son:2002ci} for any $T \lesssim T_C$, i.e.,  
\begin{align}
	\omega_{\boldsymbol{p}} = u(T)\sqrt{\boldsymbol{p}^2 + m_{\pi}^2}.
	\label{eq:disprel}
\end{align}
The associated chiral expansion around $(T\neq 0, m_q = 0)$ is based on the assumption that one is sufficiently close to the chiral limit. In this limit, the `pion velocity' $u(T)$ is the group velocity of a massless pion excitation. At finite but small quark mass and vanishing momentum it is the ratio between the quasiparticle mass $\omega_{\boldsymbol{0}}$ and the screening mass $m_{\pi}$.
For any temperature below $T_C$ we use two independent lattice estimators for the parameter $u(T)$ \cite{Brandt:2014qqa}:
\begin{align}
	u_m = \left[- \frac{4m_q^2}{m_{\pi}^2}\left.\frac{G_P(x_0,T,\boldsymbol{p}=0)}{G_A(x_0,T,\boldsymbol{p}=0)}\right|_{x_0 =\beta/2}\right]^{1/2},
	\label{eq:um}
\end{align}
\begin{align}
	u_f = \frac{f_{\pi}^2 m_{\pi}}{2 G_A(\beta/2,T,\boldsymbol{p}=0)\sinh(u_f m_{\pi}\beta/2)}. 
	\label{eq:uf}
\end{align}
We use the definition of the quark mass~$m_q$ given by the PCAC relation and $\beta$ denotes the inverse temperature.
The consistency of both estimators serves as an indicator for the applicability of the chiral expansion. Since $u(T)$ is a renormalization group invariant quantity, $u_m$ and $u_f$ also do not need any renormalization. 
The required time-dependent Euclidean correlators are  
\begin{eqnarray}
	\delta^{ab} G_A(x_0, T, \boldsymbol{p})	= \int \mathrm{d}^3 x\; e^{i \boldsymbol{p}\cdot\boldsymbol{x}} \langle A^a_0(x) A^b_0(0) \rangle =
	 \delta^{ab} \int_{0}^{\infty} \mathrm{d} \omega\; \rho_{P}\left(\omega,\boldsymbol{p}\right)\frac{\cosh(\omega(\beta/2-x_0))}{\sinh(\omega\beta/2)}, \\
	\delta^{ab} G_P(x_0, T, \boldsymbol{p})	= \int \mathrm{d}^3 x\; e^{i \boldsymbol{p}\cdot\boldsymbol{x}} \langle P^a(x) P^b(0) \rangle = \delta^{ab} \int_{0}^{\infty} \mathrm{d} \omega\; \rho_{P}\left(\omega,\boldsymbol{p}\right)\frac{\cosh(\omega(\beta/2-x_0))}{\sinh(\omega\beta/2)},
\end{eqnarray}
where $a,b$ are adjoint $\mathrm{SU}(2)_{\mathrm{isospin}}$ indices. The definition of the estimators $u_f$ and $u_m$ further contains the pion screening decay constant $f_{\pi}$ and the
screening pion mass $m_{\pi}$ which are defined by the asymptotic behavior of screening correlators, e.g., 
\begin{align}
  		  \delta^{ab} G^{s}_{A} \left(x_3, T, \boldsymbol{p}=0 \right) = \int \mathrm{d}x_0 \mathrm{d}^2 x_{\perp}
	\left\langle A^a_{3} \left(x\right) A^b_{3} \left(0\right) \right\rangle
	\overset{\left|x_3\right| \rightarrow \infty }{=}  \delta^{ab}\frac{f^{2}_{\pi} m_{\pi}}{2} e^{- m_{\pi} \left|x_3\right|}.
  \end{align}
The PCAC relation implies that at zero spatial momentum the spectral functions in P and/or A channels\footnote{See~\citep{Brandt:2014qqa, Brandt:2015sxa} for definitions in the `AP' channel.} are related by 
\begin{align}
	\rho_{AP}\left(\omega,\boldsymbol{p}=0\right) = \frac{\omega}{2 m} \rho_{A}\left(\omega,\boldsymbol{p}=0\right),
\end{align}
\begin{align}
	\rho_{P}\left(\omega,\boldsymbol{p}=0\right) = -\frac{\omega^2}{4 m^2} \rho_{A}\left(\omega,\boldsymbol{p}=0\right).
\end{align}
Therefore, in order to extract the screening quantities, we formulate a simultaneous one-state fit ansatz for the corresponding correlation functions of the form
\begin{eqnarray}
	\qquad\qquad\qquad   &G^{s}_{A} \left(x_3, T,\boldsymbol{p}=0\right) = \frac{A_1^2 m_1}{2}\cosh\left[m_1\left(x_3 - L/2\right)\right], \label{eq:simfit1}\\ 			
	\qquad\qquad\qquad   &G^{s}_{AP} \left(x_3, T,\boldsymbol{p}=0\right) = \frac{A_1^2 m^2_1}{2}\sinh\left[m_1\left(x_3 - L/2\right)\right],\\
	\qquad\qquad\qquad  &G^{s}_{P} \left(x_3, T,\boldsymbol{p}=0\right) = -\frac{A_1^2 m^3_1}{8 m^2_q}\cosh\left[m_1\left(x_3 - L/2\right)\right]. 
	\label{eq:simfit2}
\end{eqnarray}
Here we also used that $G^s_A,G^s_P$ are symmetric around $x_3 = L/2$ and that $G^s_{AP}$
is antisymmetric around $x_3 = L/2$ on the lattice. 
We choose for the fit parameters values corresponding to a small (uncorrelated) $\chi^2$/d.o.f. which are stable with respect to small variations of the fit intervals for each correlator.
\subsection{Lattice setup}
\label{sec:G8}
The calculation was performed on a newly generated $20\times 64^3$ lattice ensemble with $a=0.0658(7)(7)~\unit{fm}$ and $T/T_c\approx0.91$ relative to $T_c\approx165~\unit{MeV}$ determined in the chiral limit for two dynamical degenerate light quarks~\cite{Brandt:2013mba,Brandt:2016daq}.
The simulation uses the plaquette gauge action and the $\mathcal{O}(a)$-improved Wilson fermion action with a non-perturbatively determined $c_{\text{sw}}$ coefficient~\cite{Jansen:1998mx}. The values of the bare parameters in the lattice action amount to $\beta = 5.30$ and $\kappa= 0.13642$. The pion mass for the corresponding zero-temperature CLS ensemble (G8) takes a value of $m_{\pi}(T=0)=\omega^0_{\boldsymbol{0}} \approx 185 \unit{MeV}$~\citep{DellaMorte:2017dyu}.

\subsection{Results}
\label{sec:G8res}
The results of the analysis based on Eqs.~(\ref{eq:um}, \ref{eq:uf}, \ref{eq:simfit1}-\ref{eq:simfit2}) are shown in Table~\ref{tab1:G8}. The new ensemble confirms the  qualitative findings of our previous work. The deviation of the pion velocity $u\approx 0.77$ from unity indicates a violation of boost invariance due to the thermal medium. Both estimators $u_f$ and $u_m$ are in very good agreement signaling the validity of the chiral expansion.
The calculated pion quasiparticle mass $\omega_{\boldsymbol{0}} = u_m m_{\pi} \approx 155~\unit{MeV}$ is significantly lighter than at zero-temperature, i.e., the pole mass is shifted downward by approximately 16\%. In contrast to that the pion screening mass $m_{\pi} \approx 201~\unit{MeV}$ is increased compared to the zero-temperature pion mass. 
		\begin{table}[tbh]
		\centering
		\setlength{\tabcolsep}{1mm}
		\renewcommand{\arraystretch}{1.15}
		\begin{tabular}{lr}
		\hline
		\hline
		$m_{\pi}/T$ 										  & $1.33(4)$	\\
		$f_{\pi}/T$ 										  & $0.502(4)$\\
		\hline
		$u_f$ 		&	$0.778(30)$\\
		$u_m$ 		&	$0.776(38)$\\
		\hline 
		$\omega_{\boldsymbol{0}}/T$ & $1.03(3)$ \\
		\hline
		\hline
		\end{tabular}
		\caption{Summary of the results of the $N_{\tau} = 20$ thermal ensemble described in Sec.~\ref{sec:G8}. The pion quasiparticle mass $\omega_{\boldsymbol{0}}$ is calculated using $\omega_{\boldsymbol{0} } = u_m m_{\pi}$. For $f_{\pi}$ renormalization factors~\citep{Fritzsch:2012wq} are included.}
		\label{tab1:G8}
	\end{table}
	
\subsection{Axial-charge density correlator at $\boldsymbol{p}= 0$ and $\boldsymbol{p} \neq 0$}
\begin{figure}[h] 
  \centering
  \includegraphics[width=.5\textwidth,clip]{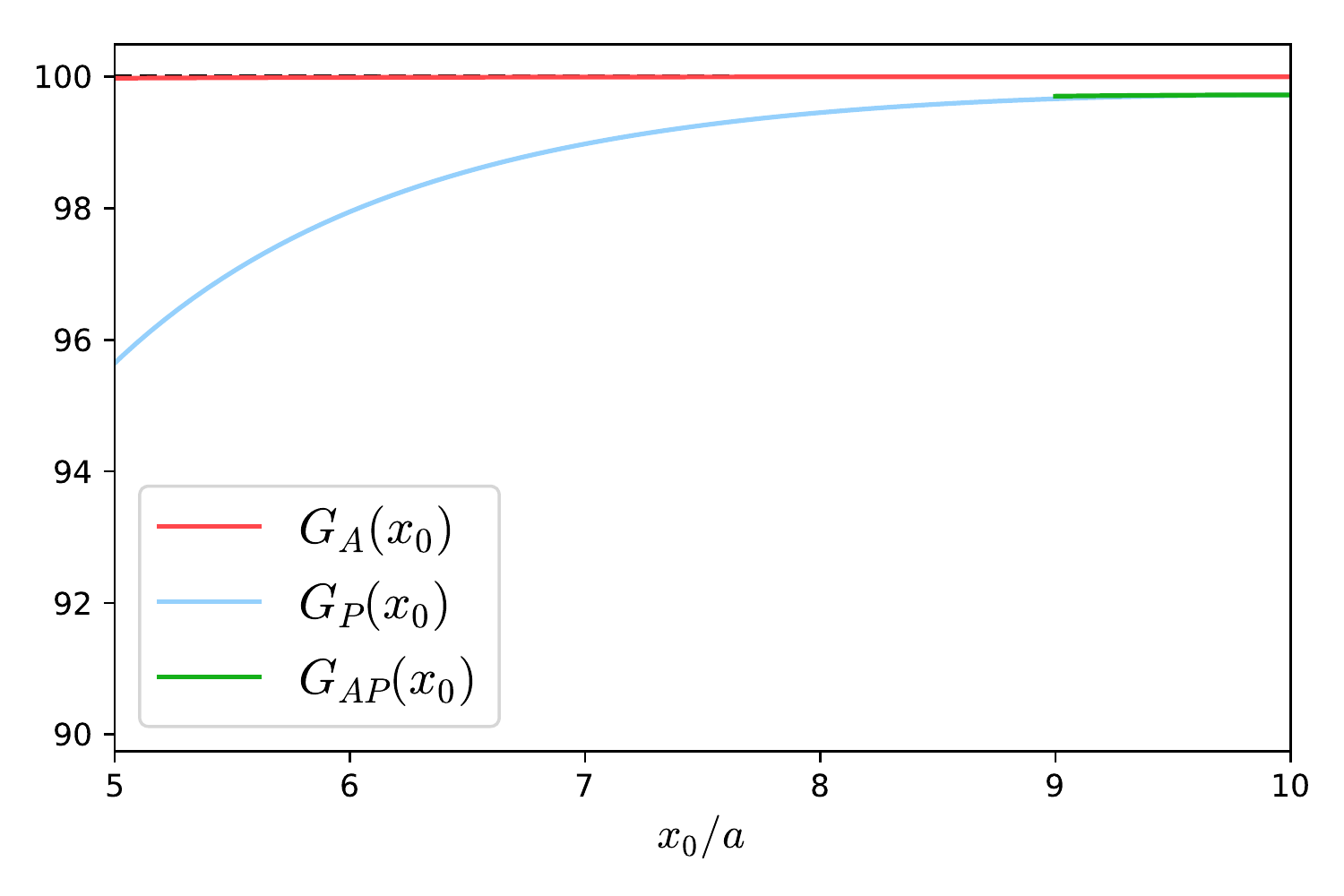}
  \caption{Pion contribution in percent for the different channels at zero momentum.}
  \label{fig-pidom}
\end{figure}

In order to test whether the parameter $u$ determined from the ratio of the quasiparticle to the screening mass really does predict the dispersion relation of the quasiparticle, as in Eq.~\ref{eq:disprel}, we analyze the time-dependent Euclidean correlator $G_A\left(x_0,T,\boldsymbol{p}\right)$. In the high frequency region, a leading-order perturbative calculation for its spectral function, see e.g. \cite{Aarts:2005hg}, yields
\begin{align}
	\rho_{A}\left(\omega,T,\boldsymbol{p}\right) = \theta(\omega^2-4m^2-\boldsymbol{p}^2)\frac{N_c}{24 \pi^2} (\boldsymbol{p}^2+6m^2), \quad \omega \rightarrow \infty.
\end{align}
It is only at sufficiently small quark masses $m$ and momenta $\boldsymbol{p} = (0,0,2\pi n/L)$, and not too small $x_0$, that the correlator is parametrically dominated by the pion pole.
Taking these non-pion contributions into account leads to the following fit ansatz for the spectral function
\begin{align}
    \rho_{A}\left(\omega,T\boldsymbol{p}\right) = A_1(\boldsymbol{p}) \omega \sinh\left(\omega \beta/2\right) \delta\left(\omega - \omega_{\mathbf{0}}\right) +
    A_2(\boldsymbol{p}) \frac{N_c}{24\pi^2}\left(1- e^{-\omega\beta}\right) \theta\left(\omega - c\right).
\end{align}
The corresponding fit ansatz for the spectral function reads 
\begin{align}
		G_A\left(x_0,T,\boldsymbol{p}\right) = A_1(\boldsymbol{p}) \omega_{\mathbf{0}} \cosh\left(\omega_{\mathbf{0}}\left(\beta/2 - x_0\right)\right) 
    + A_2(\boldsymbol{p}) \frac{N_c}{24\pi^2}
    \left(\frac{e^{-c x_0}}{x_0} + \frac{e^{-c\left(\beta-x0\right)}}{\beta - x_0} \right).
\end{align}
At finite momentum, leaving all four fit parameters free leads to poorly constrained fits. 
Therefore, the results obtained from the previous analysis of Sec.~\ref{sec:G8res} were used to fix $\omega_{\boldsymbol{p}}$ to the predicted value given by Eq.~\ref{eq:disprel}.
However, at zero momentum we can again use the PCAC relation to perform a simultaneous fit, including data points from time-dependent correlators $G_P$ and $G_{AP}$. In this case $\omega_{\boldsymbol{0}}$ can be kept free.
The fit parameter $A_1(\boldsymbol{p})$ is related to the residue of the pion pole via
\begin{align}
	\text{Res}\left(\omega_{\boldsymbol{p}}\right) = 2 A_1(\boldsymbol{p}) \omega_{\boldsymbol{p}}^2 \sinh\left(\omega_{\boldsymbol{p}}\beta/2\right).
\end{align}
To compare the fit results to the prediction from the chiral effective field theory we introduce a parameter $b\left(\boldsymbol{p}\right)$ which parametrizes a deviation from the expected residue
\begin{align}
	\text{Res}\left(\omega_{\boldsymbol{p}}\right) = f_{\pi}^2\left( m_{\pi}^2 + \boldsymbol{p}^2\right)\left(1+b\left(\boldsymbol{p}\right)\right).
\end{align}
The results for vanishing and finite momentum are shown in Table~\ref{tab:finmom}. The pion quasiparticle mass obtained from the simultaneous fits to the time-dependent correlators at zero momentum is consistent with the prediction from the previous analysis and the parameter $b$ being compatible with zero indicates the validity of the chiral prediction. Figure~\ref{fig-pidom} shows the relative pion contribution to the fits for the different channels. At the midpoint the non-pion contributions are at the sub-percent level indicating the dominance of the pion which is an essential assumption in the definition of the estimators for the pion velocity $u$.
The results at finite momentum are qualitatively in agreement with our previous findings~\citep{Brandt:2015sxa}, namely at the lowest momentum $b\left(\boldsymbol{p}\right)$ is very small as expected from the effective theory.
\begin{table}[thb]
	\renewcommand{\arraystretch}{1.15}
\begin{minipage}[tbh]{0.99\textwidth}
	\centering
		\begin{tabular}{cc|c|c|c|c|c}
		\hline
		\hline
		$n$&$\omega_{\boldsymbol{0}}/T$ & $A_2(\boldsymbol{0})/6m^2$ & $c/T$ & 	$Res(\omega_{\boldsymbol{0}})/T^4$&	
		b & $\chi^2 / d.o.f$		    \\
		\hline
		$0$ & $1.04(3)$ & $2.74(97)$ & $10(5)$ & $0.45(4)$ & $0.007(34)$  & $0.28$ \\
 		\hline
 		\hline
 		\\[-0.2cm] 
		\hline	
		\hline	
		$n$ & $\omega_p/T$ & $A_2(\boldsymbol{p})/\boldsymbol{p}^2$ & $c/T$ & $Res(\omega_{\boldsymbol{p}})/T^4$&	
		b & $\chi^2 / d.o.f$		    \\
		\hline
		$1$ & $1.84(6)$ & $6(4)$ & $11(3)$ & $1.316(99)$ & $-0.07(5)$  & $0.19$ \\
		$2$ & $3.22(11)$ & $2.12(19)$ & $8.4(6)$ & $3.6(4)$ & $-0.17(7)$  & $0.57$ \\
		$3$ & $4.69(17)$ & $1.52(14)$ & $7.98(65)$ & $1.2(6.1)$ & $-0.424(98)$  & $0.64$ \\		
 		\hline
		\hline
		\end{tabular}		
		\label{tab:finmom}
\end{minipage}
\caption{Top: Results of the fits to the time-dependent correlators at vanishing momentum. Bottom: Results of fits to the axial-charge density correlator at non-vanishing momentum $\boldsymbol{p}_n = (0,0,2\pi n/L)$. Here the quantity $\omega_p/T$ is set to the value predicted by Eq.~\ref{eq:disprel} with $u(T) = u_m = 0.77(4) $. Renormalization factors are included~\citep{Fritzsch:2012wq}.}
\end{table}

\section{Gell-Mann--Oakes--Renner relation}\label{sec-GMOR}
\begin{figure}[tbh]
	\centering
	\includegraphics[width=.6\textwidth,clip]{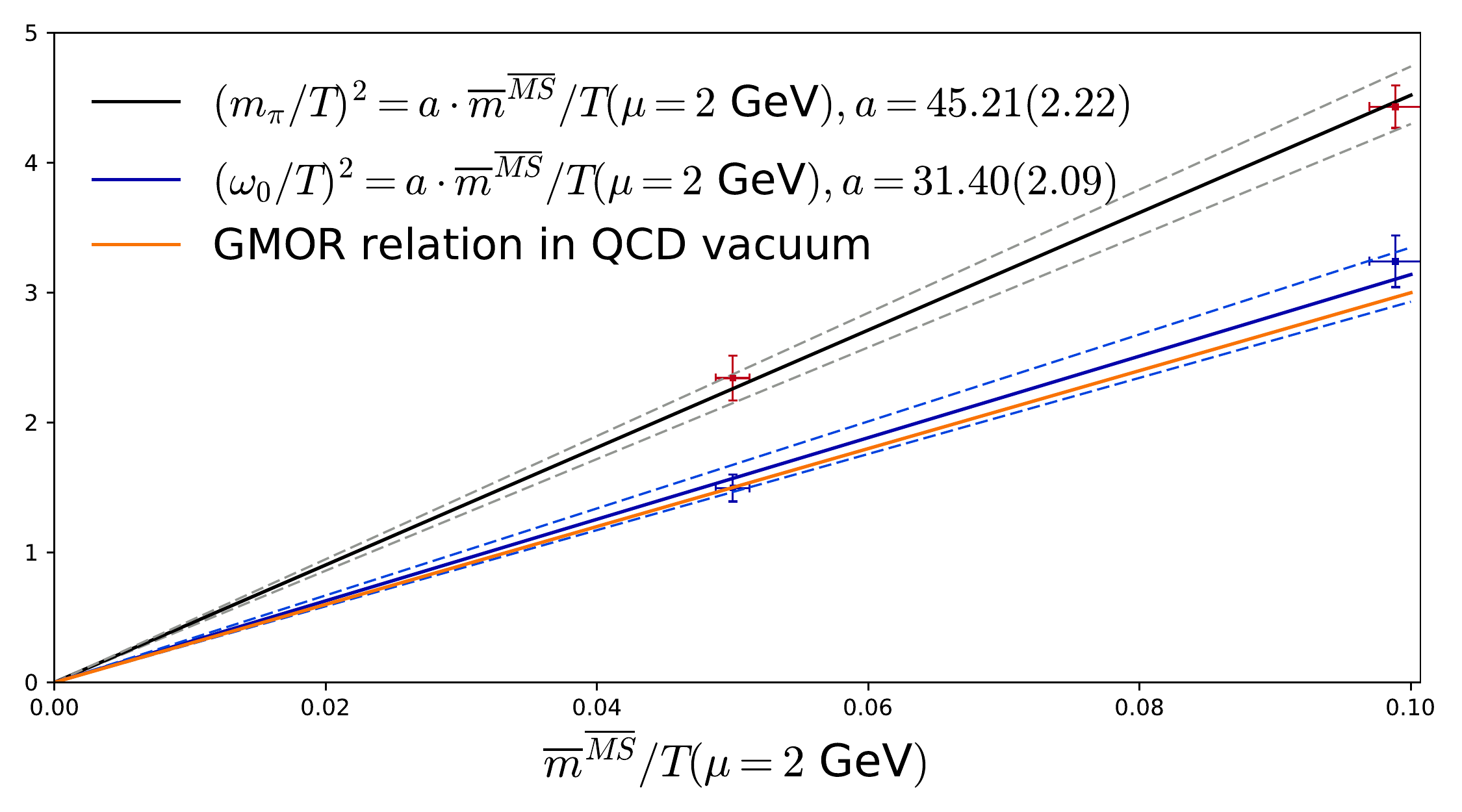}
	\caption{Test of the GMOR relation at $T\approx 150\unit{MeV}$. The vacuum result is taken from \cite{Engel:2014cka}.}
	\label{fig:GMOR}
\end{figure}

In this section we want to study the Gell-Mann--Oakes--Renner (GMOR) relation at finite temperature.
The GMOR relation gives the pion mass in terms of the pion decay constant, the chiral condensate and the renormalized quark mass $m^{(r)}$
\begin{align}
	m_{\pi}^2 = - \frac{\left\langle \overline{\psi}\psi \right\rangle}{f^2_{\pi}} \cdot m^{(r)}. 
\end{align}
At finite temperature the pion mass splits into a pion quasiparticle mass and a screening mass. We have calculated both on two different ensembles with different quark masses at fixed inverse coupling $\beta$ and temperature $T$. The lattices are of size $16\times 48^3$ and have a lattice 
spacing of approximately $0.08~\unit{fm}$. Lattice parameters and the results based on the analysis described in Sec.~\ref{sec-Disp} are shown in Table~\ref{tab:GMOR}. Figure~\ref{fig:GMOR} shows a linear dependence of both the squared pion screening mass and the squared pion quasiparticle mass on the renormalized quark mass at a fixed temperature~$T\approx 150~\unit{MeV}$. Compared to the vacuum result the slope is much steeper for the screening quantity. This is due to the decrease of the screening pion decay constant $f_{\pi}$.
From a linear fit with zero intercept the slope can be estimated in the chiral limit, i.e., 
\begin{align}
	-\left.\frac{\left<\overline{\psi}\psi\right>}{f_{\pi}^2 T}\right|_{m=0} = 45.21(2.22)
\end{align}
In case of the pion quasiparticle, we do not observe a statstically significant thermal change compared to the zero-temperature result. The determined slope is 
\begin{align}
	-\left.\frac{\left<\overline{\psi}\psi\right>}{(f^{t}_{\pi})^2 T}\right|_{m=0} = 31.4(2.22),
\end{align}
where $f^t_{\pi} \equiv f_{\pi}/u$ denotes the pion quasiparticle decay constant. For the heavier ensemble, within uncertainties its value does not deviate from the pion screening decay constant at zero-temperature obtained from the CLS ensemble with the same bare parameters (A5
) ~\citep{Brandt:2014qqa}, i.e., $f^{t}_{\pi} / f^{0}_{\pi} = 1.010(44)$. This indicates that there is also no significant change of the chiral condensate, and therefore also no significant sign of chiral symmetry restoration.
\begin{table}
\begin{minipage}[tbh]{0.45\textwidth}
		\flushright
		\setlength{\tabcolsep}{1mm}
		\renewcommand{\arraystretch}{1.25}
		\begin{tabular}{lr}
		\hline
		\hline
		$\beta$ & $5.20$ 			    \\
		$\kappa$& $0.13594$  			\\
		$T$		& $150~\unit{MeV}$ 		\\
		\hline
		$m^{\overline{MS}}/T(\mu = 2~\unit{ GeV})$ &	$0.0989(19)$\\
		$m_{\pi}/T$ 										  & $2.105(38)$	\\
		$f_{\pi}/T$ 										  &	$0.564(8)$\\
		\hline
		$u_f$ 		&	$0.870(15)$\\
		$u_m$ 		&	$0.864(21)$\\
		\hline
		\hline
		\end{tabular}
\end{minipage}\hfill
\begin{minipage}[b]{0.45\textwidth}
		\flushleft
		\setlength{\tabcolsep}{1mm}
		\renewcommand{\arraystretch}{1.25}
		\begin{tabular}{lr}
		\hline
		\hline
		$\beta$ & $5.20$ 			    \\
		$\kappa$ & $0.13599$ \\
		$T$		& $150~\unit{MeV}$ 		\\
		\hline
		$m^{\overline{MS}}/T(\mu = 2~\unit{ GeV})$ &	$0.0500(12)$\\
		$m_{\pi}/T$ 										  & $1.53(6)$	\\
		$f_{\pi}/T$ 										  &	$0.490(15)$\\
		\hline
		$u_f$ 		&	$0.801(32)$\\
		$u_m$ 		&	$0.799(41)$\\
		\hline
		\hline
		\end{tabular}
\end{minipage}\hfill
\caption{Ensembles of size $16\times 48^3$ for the test of the GMOR relation. The quark mass  is renormalized in the $\overline{MS}$ scheme at a scale of $\mu=2~\unit{GeV}$. Renormalization factors are taken from~\citep{Fritzsch:2012wq}.}
\label{tab:GMOR}
\end{table}

\section{Dey-Eletsky-Ioffe mixing theorem at finite quark mass}\label{sec-Ioffe}
In the chiral limit, the heat bath in the low-temperature phase is dominated by massless pions. Taking only the lowest states into account, 
namely the vacuum and single pion states, it was shown in \cite{Dey:1990ba} that to order $T^2$ the finite-temperature vector and axial-vector correlators can be described as a mixture between their vacuum counterparts. In terms of the corresponding spectral functions this statement reads
\begin{align}
		\rho_{V} (\omega,\boldsymbol{p},T) = (1-\epsilon)\rho_V(\omega,\boldsymbol{p},T=0) + \epsilon \rho_{A} (\omega,\boldsymbol{p},T=0), \\
		\rho_{A} (\omega,\boldsymbol{p},T) = (1-\epsilon)\rho_A(\omega,\boldsymbol{p},T=0) + \epsilon \rho_{V} (\omega,\boldsymbol{p},T=0),
\end{align}
where $\epsilon \equiv T^2/6f^2_{\pi}$ is a temperature dependent coefficient and $f_{\pi} = 93\unit{MeV}$ denotes the pion decay constant (at $T=0$). In particular, this implies that the difference of the vector and the axial spectal function at finite-temperature is proportional to its zero-temperature equivalent:
\begin{align}
			\rho_{V} (\omega,\boldsymbol{p},T) - \rho_{A} (\omega,\boldsymbol{p},T) = 
			(1-2\epsilon)
			\left[\rho_{V} (\omega,\boldsymbol{p},T=0) - \rho_{A} (\omega,\boldsymbol{p},T=0)\right].
			\label{eq:DEI}
\end{align}
Since this quantity is an order paramter for chiral symmetry restoration, it is instructive to investigate its behavior even at a finite quark mass. We consider the difference `V-A' of the corresponding correlators at zero momentum
\begin{align}
	\delta^{ab} \left[ G_V(x_0, T, \boldsymbol{p} = 0) - G_A(x_0, T, \boldsymbol{p} = 0) \right]	= \int \mathrm{d}^3 x\; \sum^3_{i=1}\left[ \langle V^a_i(x) V^b_i(0) \rangle - \langle A^a_i(x) A^b_i(0) \rangle\right].
\end{align}

\begin{figure}[tbh]
	\begin{minipage}[t]{0.49\linewidth}
		\includegraphics[width=0.99\textwidth]{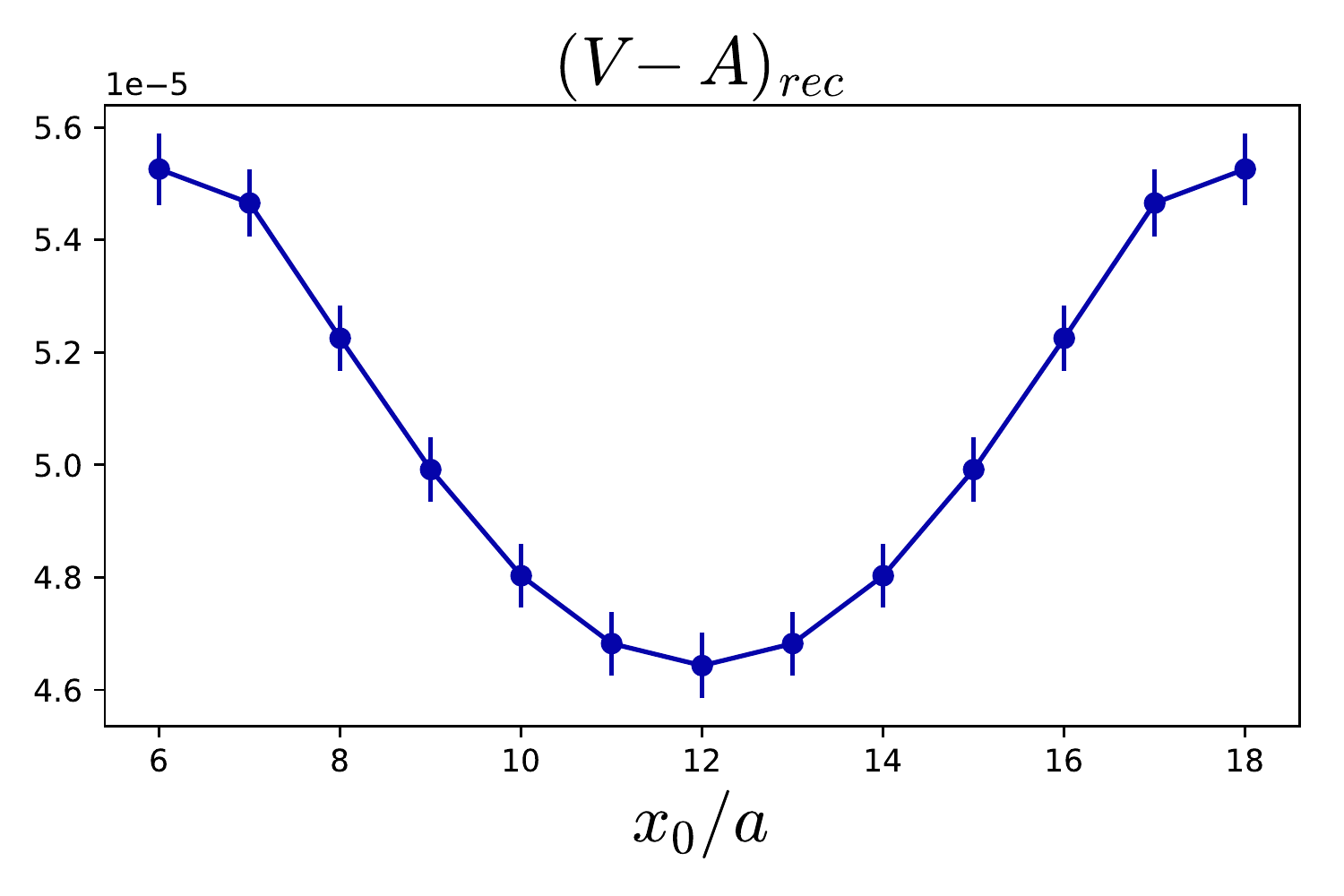}
	\end{minipage}
	\begin{minipage}[t]{0.49\linewidth}
		\includegraphics[width=0.99\textwidth]{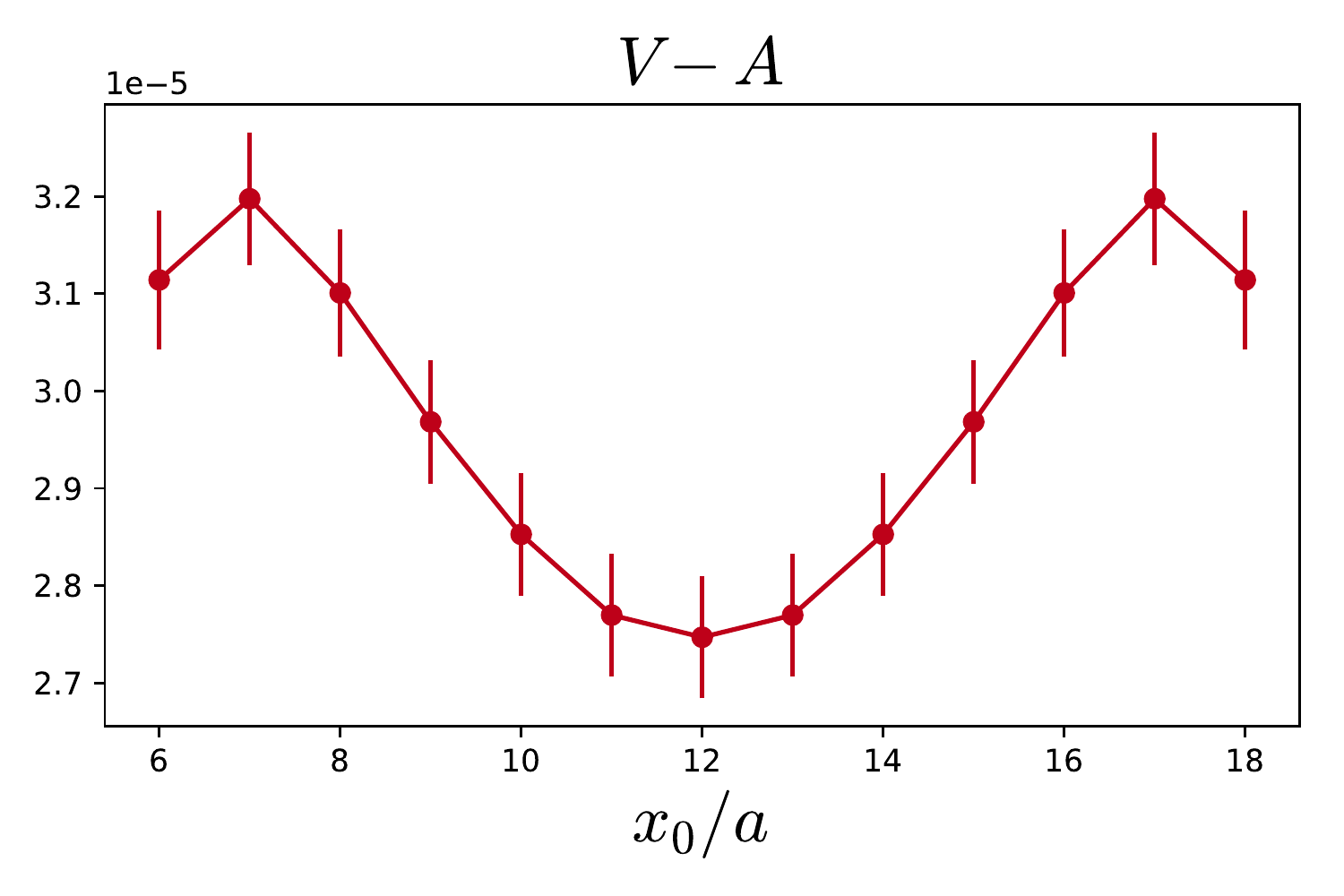}
	\end{minipage}
	 \caption{Left: The reconstructed correlator for the difference '$V-A$'. Right: The difference of `$V-A$' at $T\approx 169~\unit{MeV}$. All renormalization factors are included~\cite{Fritzsch:2012wq, DellaMorte:2005xgj}.}
  \label{fig-1}
\end{figure}

Measurements are performed on a $24\times 64^3$ lattice with two dynamical light quarks with a mass of $ m^{\overline{MS}}(\mu = 2\unit{GeV}) = 12.8(1)~\unit{MeV}$. The temperature is $T = 1/24a \approx 169\unit{MeV}$. Furthermore, we use a $128\times 64^3$ corresponding quasi zero-temperature CLS ensemble (O7) to obtain a comparable effectively zero-temperature quantity
for the difference `$V-A$'\footnote{See \cite{Brandt:2015sxa} for more details on the ensembles. }. This is achieved by calculating the `reconstructed' correlator for the difference, i.e., the thermal Euclidean correlator that would be realized if the spectral function was unaffected by thermal effects ~\cite{Meyer:2010ii}.
Figure~\ref{fig-1} shows the reconstructed correlator for the difference `$V-A$' and the
same quantity for the thermal ensemble.

\begin{figure}[thb] 
  \centering
  \includegraphics[width=.55\textwidth,clip]{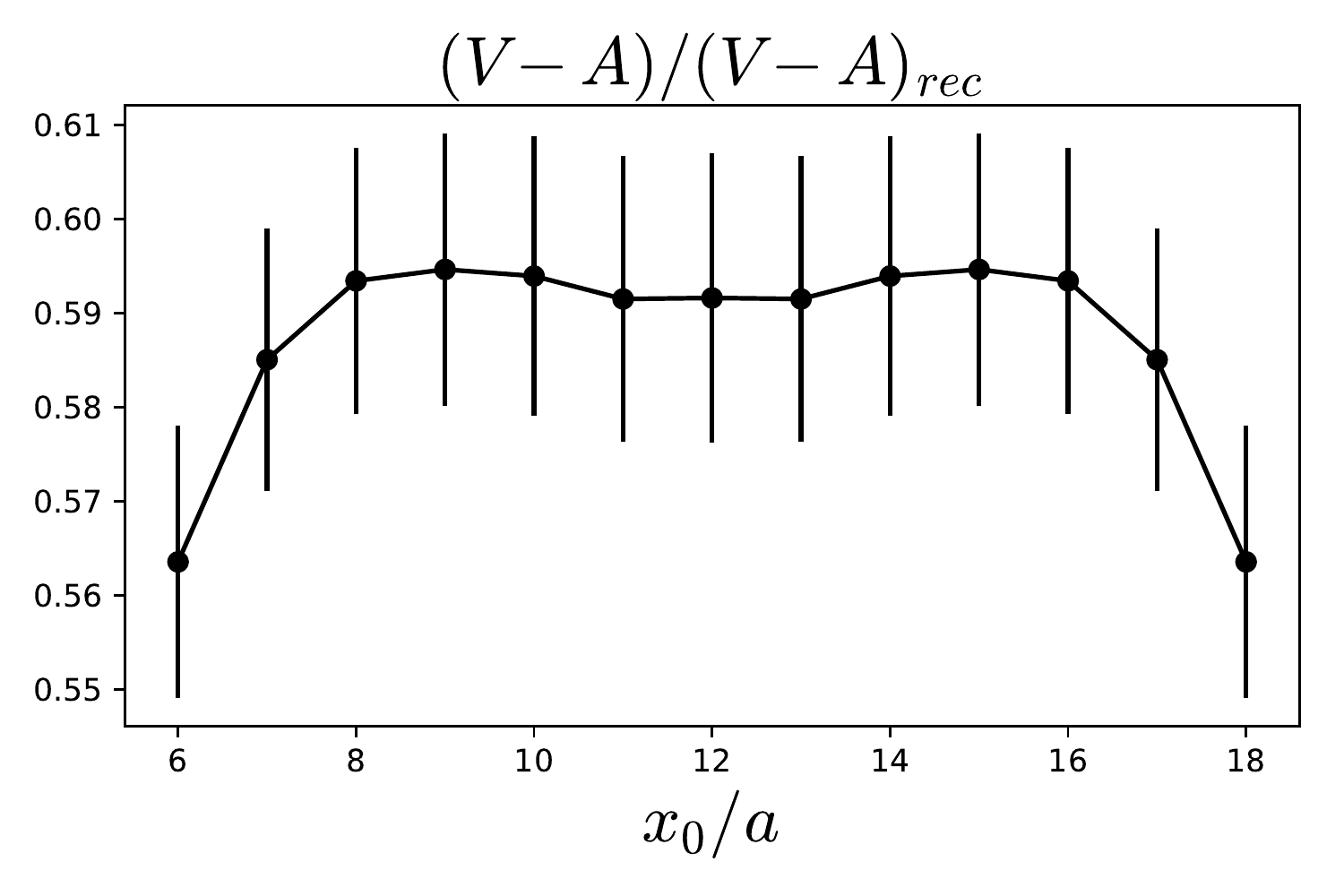}
 	\caption{Ratio of the difference '$V-A$' and the reconstructed correlator.}
  \label{fig-2}
\end{figure}

Their ratio is shown in Figure~\ref{fig-2}. Even for a finite quark mass of approximately $12.8~\unit{MeV}$ it is very flat, consistent with the prediction of Eq.~\ref{eq:DEI} obtained in the chiral limit. The difference `$V-A$' shows a significant reduction, by a factor of approximately $0.6$ at $T\approx 169\unit{MeV}$. Therefore, chiral restoration is at an advanced stage in the spectral function. The interested reader is
also referred to \cite{Brandt:2016daq}, where a rapid approach to chiral restoration in the difference of vector and axial-vector screening masses was observed. 
\section{Conclusions and outlook}
We have investigated the pion quasiparticle and order parameters
for chiral restoration in $N_f=2$ QCD. The results obtained from the new ensemble with a lighter quark mass are qualitatively in agreement with our previous findings \citep{Brandt:2014qqa, Brandt:2015sxa}, namely that the pion mass splits in a significantly lighter pion quasiparticle mass and a heavier screening mass. Furthermore, we
successfully tested that the modified dispersion relation given in Eq.~\ref{eq:disprel} is determined by the `pion velocity' $u\approx 0.77$ corresponding to a violation of boost invariance due to the thermal medium. 
We find no deviation from the GMOR relation up to the considered quark mass $ m^{\overline{MS}}(\mu = 2\unit{GeV}) \approx15~\unit{MeV}$. In addition, two different order parameters have been investigated. The chiral condensate obtained from the GMOR relation shows no significant thermal change at $T\approx 150~\unit{MeV}$ indicating no sign of chiral symmetry restoration. In contrast to that, at $T\approx 169~\unit{MeV}$, the `$V-A$' correlator is notably reduced by a factor of approximately $0.6$ compared to its value at zero-temperature. Here, chiral restoration in the `$V-A$' spectral function is at a more advanced stage.

Future considerations might include the possibility of a finite width of the pion quasiparticle. Additionally, for a more rigorous comparison of the pion screening mass it might be instructive to extract the equivalent quantity from corresponding zero-temperature spatial correlation functions.
Furthermore, we plan to extend our studies to $N_f = 2+1$, including a dynamical strange quark.
\section{Acknowledgements}
We acknowledge the use of computing time on the JUGENE and JUQUEEN
computers of the Gauss Centre for Supercomputing located at
Forschungszentrum Juelich, Germany under grant HMZ21. Part of the
calculations were performed on the cluster `Clover' of the
Helmholtz-Institute Mainz. This work was supported by the DFG Grant
No. ME 3622/2-2 \emph{QCD at non-zero temperature with Wilson fermions
on fine lattices}.
\footnotesize
\bibliography{lattice2017}

\end{document}